# *Prognosis of Anterior Cruciate Ligament (ACL) Reconstruction: A Data Driven Approach*

*Abhijit Chandra[1], Oliva Kar[2], Kuan-Chuen Wu[1], Michelle Hall[3] and Jason Gillette[3]*
[1]Mechanical Engineering
[2]Computer Science and Human Computer Interaction
[3]Kinesiology
Iowa State University
Ames, Iowa
Contact: achandra@iastate.edu

*Abstract* — **Individuals who suffer anterior cruciate ligament (ACL) injury are at higher risk of developing knee osteoarthritis (OA) and almost 50% display symptoms 10-20 years post injury.** *Anterior cruciate ligament reconstruction (ACLR) often does not protect against knee OA development.* **Accordingly, a multi-scale formulation for Data Driven Prognosis (DDP) of post-ACLR is developed. Unlike traditional predictive strategies that require controlled off-line measurements or "training" for determination of constitutive parameters to derive the transitional statistics, the proposed DDP algorithm relies solely on in-situ measurements. The proposed DDP scheme is capable of predicting onset of instabilities. Since the need for off-line testing (or training) is obviated, it can be easily implemented for ACLR, where such controlled a priori testing is almost impossible to conduct. The DDP algorithm facilitates hierarchical handling of the large data set, and can assess the state of recovery in post-ACLR conditions based on data collected from stair ascent and descent exercises of subjects. The DDP algorithm identifies inefficient knee varus motion and knee rotation as primary difficulties experienced by some of the post-ACLR population. In such cases, levels of energy dissipation rate at the knee, and its fluctuation may be used as measures for assessing progress after ACL reconstruction.**

## I. INTRODUCTION

Individuals who suffer an anterior cruciate ligament injury are at a higher risk of developing knee osteoarthritis (OA) and almost 50% display symptoms 10-20 years post injury [1]. Anterior cruciate ligament reconstruction (ACLR) often results in return to play for young athletes, but does not protect against knee OA development. Therefore, a young post-ACLR population (25-40 years old) is at increased risk for the early onset of chronic pain and functional impairments associated with knee OA. Those with knee OA experience a dramatic decline in their physical activity level, with 40.1% of men and 56.5% of women considered inactive [2].

Proposed mechanisms for early onset knee OA include increased knee joint loading and/or unfavorable changes in neuromuscular control that develop post-ACLR. A 1% increase in external knee varus moment is estimated to increase the risk of OA progression by a factor of six [3]. However, studies have been equivocal as to whether external knee varus moments are increased, unchanged, or decreased post-ACLR [4-5]. Increases in quadriceps-hamstring-gastrocnemius co-contraction have been correlated with increased knee OA severity [6]. Co-contraction may act to stabilize the knee post-ACLR, but may also increase knee joint compression and/or negatively shift loading to the medial knee joint.

Our preliminary research has focused on comparing individuals post-ACLR and apparently healthy controls during walking, stair ascent, and stair descent. The post-ACLR group demonstrated lower knee extensor moments and unchanged external knee varus moments, but higher quadriceps-hamstrings co-contraction [5, 7]. Therefore, maximum joint moments may not be sensitive enough to detect early changes in knee joint loading until knee OA is in an advanced state. Multi-physics modeling examines the energy balance between knee joint moments and whole body movement. In particular the level of energy flow through the knee as well as its fluctuation is studied. Energy dissipation in the knee joint may lead to breakdown of articular cartilage and provide early detection for the development of knee OA. It is interesting to note that large energy dissipation rate results in larger internal forces experienced by the knee that may cause increased wear and large varus or rotational motion with progression of time.

Predictive or prognostic capability represents a core competency required in many walks of life. Traditionally, analytical or numerical models based on conservation principles are used to make such predictions. These models require satisfaction of three principles: (1) equilibrium (which embodies satisfaction of a conservation principle), (2) compatibility, and (3) constitutive relations. In most cases, the constitutive relations depend on prior knowledge of constitutive or material parameters. Most often, a failure criterion and associated critical or threshold value of a parameter (typically a scalar) is also required to predict onset of instabilities. Such estimation protocols require controlled off-line testing or "training" for the prediction model. Further, applicability of such a prediction model is limited to situations spanned by the "column space" of the training set.

Numerous applications of practical interest also suffer from two major drawbacks: (1) the explicit expression for conservation principle or utility function being conserved may be unknown, (2) the system may not be amenable to perform off-line (in particular, destructive) testing. For instance while trying to predict the onset of instability in a material (e.g. necking in a tension test done on a material), the material properties can be estimated even before the actual experiment is done. The inferred constitutive parameters are then utilized for prognosis during the actual experiment. But if the part under consideration is already in service (e.g., in a structure or part of a person's anatomy, e.g., knee or ACL), it may not be accessible for such off-line parameter evaluation. This is also true when a non-materialistic or abstract system (e.g. a global economic system, genetics or a health care system) is considered, there is neither any way to know the specific system properties, nor there is any method to assess the explicit nature of the conservation potential. In such cases, there is a need to rely on a predictive algorithm that can function without a priori knowledge of these parameters.

The present work circumvents these two difficulties by devising a DDP algorithm. The proposed algorithm requires assumption of the existence of a conservation principle, but its exact form need not be specified a priori. Instead, the exact form of the conservation functional is only specified locally, in the neighborhood of each observation point, (and is assumed to be piecewise quadratic in the current work) while its global form remains unknown. The conservation principle is then defined as the minimization of system curvature at each of the observation points. Based on such minimization principle, a dimensionless length scale of the underlying phenomenon is estimated at each observation point in each individual dimension. The constitutive parameters of any system (physical or abstract) are contained (with correct combination rule) within the expression for dimensionless length scales. Finally, stability is defined as the ability of the system to minimize its local curvature below a threshold value (determined as inverse of length scale in the present work) at each individual observation points. Thus, the stability characteristic of a system at a given location depends only on the local curvature of the system and the instantaneous dimensionless length scale at that location, obviating the need for explicit estimation of constitutive parameters.

## II. LITERATURE REVIEW

The key to useful prognosis is consistent estimation of future trend based on past experience and current observations defining the current state of the system. Two key questions are: (1) Will the current trend continue over a pre-specified prediction horizon? And (2) What is the remaining life or time-span over which current activities can be continued? A detailed literature review along these lines may be found in [20].

Prognostics and health management (PHM) capabilities combine sensing and interpretation of data on the environmental, operational, and performance-related parameters of the product to assess the health of the product and then predict remaining useful life (RUL). This data is often collected in real-time or near real-time and used in conjunction with prediction models to provide an estimate of its state-of-health or degradation and the projection of remaining life. Traditionally, these prediction models implement either a model based approach or a data-driven approach.

Model based approaches are based on an understanding of the physical processes and interrelationships among the different components or subsystems of a product [21], including system modeling and physics-of-failure (PoF) modeling approaches.

In system modeling approaches, mathematical functions or mappings, such as differential equations, are used to represent the product. The constitutive parameters or coefficients of the differential operators must be known a priori or estimated off-line based on a controlled set of stimulations. Statistical estimation techniques based on residuals and parity relations are then used to detect, isolate, and predict degradation [21-22]. Model-based prognostic methods have been developed for digital electronics components and systems such as lithium ion batteries [23], microprocessors in avionics [12], global positioning systems [13], and switched mode power supplies [23].

One of the examples of statistical estimation techniques and system modeling that is utilized for anomaly detection, prediction and diagnosis is the "divide-and-conquer" [21] dynamic modeling paradigm. The system input–output operation space is partitioned into small regions using self-organizing maps (SOMs) and then a statistical model of the system expected behavior within each region is constructed based on time–frequency distribution (TFD).The significant deviations from the trained normal behavior are recognized as anomalies. Then "diagnosers" can be constructed for various known faults within each operational region to identify the types of faults. This divide-and-conquer approach leads to a localized decision-making scheme, where anomaly detection and fault diagnosis can be performed locally within each operational region.

Another system modeling prognostic method used for diagnosis for lithium ion batteries is the Bayesian Framework approach [23]. The Bayesian learning framework attempts to explicitly incorporate and propagate uncertainty in battery aging models. The relevance vector machine (RVM)–particle filter (PF) approach provides a probability density function (PDF) for the end-of-life (EOL) of the battery. RVM is a Bayesian form representing a generalized linear model of identical functional form of the Support Vector Machine (SVM). Both of the above methods use some sort of supervised learning.

PoF based prognostic methods utilize knowledge of a product's life cycle loading conditions, geometry, material properties, and failure mechanisms to estimate its RUL [12-15]. PoF methodology is based on the identification of potential failure mechanisms and failure sites of a product. A failure mechanism is described by the relationship between the in situ monitored stresses and variability at potential failure sites. PoF based prognostics permit the assessment and prediction of a product's reliability under its actual application conditions. It integrates in situ monitored data from sensor systems with models that enable identification of the deviation or degradation of a product from an expected normal condition and the prediction of the future state of reliability. Such methodology requires establishment of "benchmark" parameters or normal behavior of the product. This requires controlled off-line testing or controlled "training" regimen.

PoF based approach has been applied to analyze the health of printed circuit boards (PCB) to vibration loading in terms of bending curvature [12]. First, the components which are most likely to fail and their locations are identified at certain vibration loading levels. Sensors are placed at those areas to monitor the PCB response. Then a database is built that reflects the relation between the PCB and its critical components. Similar approaches have also been applied for surface mount assemblies to extract the state of damage at an instant and predict the RUL based on accumulated damage [16].

The data-driven approach uses statistical pattern recognition and machine learning to detect changes in parameter data, isolate faults, and estimate the RUL of a product [20-23]. Data-driven methods do not require product-specific knowledge of such things as material properties, constructions, and failure mechanisms. In data-driven approaches, in-situ monitoring of environmental and operational parameters of the product is carried out, and the complex relationships and trends available in the data can be captured without the need for specific failure models. There are many data-driven approaches, such as neural networks (NNs), SVMs, decision tree classifiers, principle component analysis (PCA), PF, and fuzzy logic [20]. However, such techniques also require a definition of normal operation, which is typically based on a training set or previously observed circumstances. Thus, current state-of-the-art predictions in data driven approaches only extend to circumstances that can be spanned by the 'column space' of the training conditions. This severely limits their usefulness, particularly in unforeseen circumstances. Alternatively, developing a data driven scheme for a specific purpose requires very carefully articulated training regimen (capturing the purpose) for the prognostic system.

Neural networks are examples of traditional data driven systems, where the data processing is carried out at a number of interconnected processing elements called neurons. The neurons are usually organized in a sequence of layers, an input layer, a set of intermediate layers, and an output layer. During training, the network weights are adjusted depending on the type of learning process. A set of feed forward back propagation networks are used that undergo supervised learning to identify the current operating time of an operating bearing. Two classes of neural network models, single-bearing models and clustered-bearing models have been developed [24]. Both classes of models use degradation information associated with the defective phase of bearing degradation. In first class, a single bearing is used to train a single back propagation neural network. In the second class, the bearings are classified in groups (clusters) based on similarity in their failure and defect times. Each net is then trained using degradation information associated with bearings in the cluster [24].

Thus, current state-of-the-art in both system modeling and data driven approaches is limited by the prior training requirement. Controlled a priori testing is needed for accurate model parameter estimation. Moreover, the applicability of a prognostic system is strictly limited to the vector space spanned by basis vectors of its training regimen.

By contrast, the proposed DDP approach estimates the relevant constitutive parameters in-situ. Instead of carefully designed "training" regimen, it can utilize any two time sequences of data. However, it only focuses on the situation at hand, rather than trying to master any general situation. Accordingly, only a relevant combination of material and geometric parameters is extracted as an instantaneous dimensionless length scale in each dimension, and utilized in real time. Such a DDP scheme is then capable of predicting the probability of the onset of instability over a prediction horizon. Alternatively, the prediction horizon needed for the probability of instability to exceed a threshold value can be considered the RUL. A companion paper verifies the proposed DDP algorithm against experimental observations of balloon burst under laboratory settings [33].

## III. DETERMINATION OF LENGTH SCALE

Let us consider an observable body or a phenomenon containing a finite number of observation points. At each point, information is collected at multiple dimensions (that collectively satisfy work conjugacy requirements), and at discrete instants of time. Let us now consider two specific observation points A and B. At both of these points, each dimension is denoted by $i = 1, n$. The value recorded at these two points in some particular dimension $i = d$ can be denoted as $u_d^A$ and $u_d^B$. For the present analysis, after all values are recorded at every observation point at a time instant, a normalized relationship is developed between each pair of points in each dimension. Such a normalized relationship between two points A and B is given by $a_d^{AB} = \frac{u_d^A - u_d^B}{u_d^A + u_d^B + 2\overline{m}}$, where $\overline{m}$ is an arbitrary small constant that is determined later.

In order to develop a model describing such a phenomenon, it is first assumed that the system under observation is conservative. As a first attempt, it is also assumed that a piecewise second order potential is sufficient to describe the pairwise interactions in the system. Thus, the general potential function is assumed to be quadratic in the neighborhood of each observation points. However, the nature of the quadratic potential function (coefficients of the second order polynomial) can vary from one point to the other, representing a higher order global relationship. The approximations and intrinsic uncertainty introduced in the predictions due to this approximation will be examined later.

A key conceptual difficulty is faced in the present work. Assumption of conservation in the system requires observation points to be interacting with each other. In a previous work [33], these observation points were spatially distributed in the balloon under observation. Unfortunately, the subject population under observation here, is clearly a non-interacting set. The difficulty is resolved by assuming individual subjects as "conservative and isolated systems" in themselves. Thus, a snippet of time sequences of acquired data on individual subjects is considered as a "data-burst" frame in instantaneous time, and assumed to be a fully interactive set. Such an assumption artificially exaggerates the level of interaction in the "frame" because, in reality, a data point can only influence or interact in future, but not in the past. Implications of such over-estimation remain to be investigated.

Next, it is attempted to satisfy the three canonical requirements: (1) compatibility, (2) equilibrium, and (3) constitutive relation. It is further acknowledged that objectivity or frame invariance (with respect to the observer) is a requirement for describing the behavior of such a system. Objectivity implies that the state of the observed system remains invariant with respect to different observers or variations in the observation procedure. In the present development, compatibility is enforced indirectly by requiring that the system be objective at every observable scale at each location [25]. Such an indirect approach satisfies compatibility as well as objectivity simultaneously. Using such an approach [25], the conservation of linear momentum around point A may be described as,

$$R_i^A - \beta_{ik}^A * \Delta H_k^A = 0 \qquad (1)$$

Here $R_i^A$ is the rank at point A in dimension $i$ [25]. Such a rank satisfies both compatibility (and objectivity) as well as equilibrium at A. $H_k^A$ represents the Borda Count [26-30], and $\Delta H_k^A$ represents the change in the Borda count at point A in dimension $k$ during a time step. The parameter $\beta_{ik}^A$ may be described as $\beta_{ik}^A = \frac{1}{(\Delta t)^2} * \frac{\rho}{E_{ijkl}} * \|L_l L_j\|$. $\rho$ is density and $E_{ijkl}$ is tangent modulus. The parameter $\beta_{ik}^A$ is a non-dimensional quantity and essentially represents a second order norm of the length-scale around the point A, in which the linearized (1) is valid. The length-scale essentially denotes a region around the observation point, in which the piecewise quadratic assumption for the potential function is valid. So substituting $\beta_{ik}^A$ in the equation above, we obtain an expression for length scale $(L_l)$ :

$$R_i^A - \frac{1}{(\Delta t)^2} * \frac{\rho}{E_{ijkl}} * L_l L_j * \Delta H_k^A = 0 \qquad (2)$$

$$\Rightarrow R_i^A = \frac{1}{(\Delta t)^2} * \frac{\rho}{E_{ijkl}} * L_l L_j * \Delta H_k^A \qquad (3)$$

$$\Rightarrow E_{ijkl} \left[\frac{R_i}{L_l L_j}\right] = \frac{\rho}{(\Delta t)^2} * \Delta H_k \qquad (4)$$

Out of all the possible transformation laws, the ones that are permissible in the above (4) are the ones that conserve angular momentum and the symmetry of the potential function. Now the conservation of angular momentum requires that $(i\ and\ j)$ be interchangeable. Similarly, the interchangeability of $(k\ and\ l)$ is mandated by the symmetry requirements on the definition of strain. The requirement of work conjugacy necessitates symmetry in potential function and this enforces interchangeability betweenthe $(i\ ;\ j)$ pair and the $(k\ ;\ l)$ pair. After all such transformations, the above (4) can be written as:-

$$\frac{E_{ijkl}}{2} * \left[\frac{R_i}{L_l L_j} + \frac{R_j}{L_k L_i} + \frac{R_k}{L_j L_l} + \frac{R_l}{L_i L_k}\right] = \frac{\rho}{(\Delta t)^2} * [\Delta H_i + \Delta H_j + \Delta H_k + \Delta H_l] \qquad (5)$$

Since the number of solutions for the quadratic equation in $L_i$ is $2^{dimSize}$, 8 and 16 values of $L_i$ are obtained when number of total dimensions are 3 and 4, respectively. The number of possible solutions is called the number of roots of the length scale. It is interesting to note that solution for a dimensionless form $\bar{L}$ of the length scale containing the correct combination of geometric and material parameters is sufficient for our intended prognosis.

A constant $m_i^{AB}$ is evaluated [31] for each observation pair by assuming an order (quadratic in present work) of the interaction potential between the pair under consideration. Note, this order need not be the same for all pairs. Next, the constant $m_i^{AB}$ is evaluated by setting the spatial gradient of the observed variable at observation point A $(u_{i,j}^A)$ to be exactly the same as the spatial gradient of the normalized variable $(a_{i,j}^A)$. Finally, the constant $\overline{m}$ is evaluated from a least square fit of all the $m_i^{AB}$ values. The constant $\overline{m}$ essentially sets the datum, and the coordinate of the origin is set at $-\overline{m}$ in the respective dimension. Since only a least square approximation is used for evaluating $\overline{m}$, and it is used universally at all points (to set same datum for all observation points), the gradient of the observed variable and that of the normalized variable are not exactly the same at all points. This introduces an approximation in our formulation, and decides the limiting resolution for the current DDP methodology. The error introduced due to this approximation is estimated in the present work.

Prognosis of ACL Reconstruction: A Data Driven Approach

*A. Energy Exchange Rate and Its Manifestation as a Curvature*

It is assumed that equilibrium in the system is satisfied instantly, while satisfaction of compatibility (in an objective framework) only happens with time. Thus, at every instant, the system state adapts in an attempt to satisfy compatibility in addition to equilibrium. As a result, the Borda Count at an observation point A changes over the associated length scale (which itself can also change as a result of this adaptation). The local curvature reflects this change, which is manifested as a local energy exchange (absorption or release) rate. We assume that loss of stability occurs locally when the curvature at a point A exceeds a threshold value (denoted by inverse of dimensionless length scale). It is further assumed that global instability occurs when such unstable points can form a "chain" or an energy exchange pathway, and it meets two additional conditions: (i) the chain length exceed a critical threshold, and (ii) the energy exchange rate along such a pathway exceeds a critical threshold [31].

## IV. ALGORITHMS

The algorithms needed for evaluation of length scales, curvatures as well as instability prediction criteria are discussed in [31, 33].

The "zoom out" or aggregation procedure for calculating the critical chain length and the residual (or aggregated) curvature (RC) for the system at a time instant needs a special mention here. The residual curvature provides a measure of the energy exchange rate of the system as a whole with its environment. Further details may be found in [31].

The "zoom out" procedure utilizes the fact that for a conservative and isolated system, the observed curvature should be zero when the "stand-off" distance of the observer is both zero and infinity. The x-axis in Fig. 1 represents such stand-off distance or "zoom-out" level, while y-axis represents the system curvature estimated under such situations. A log-scale in x is used for convenience. It is assumed that at x-value of zero, the system curvature drops to zero. However, due to resolution limitations of the sensing system, it was not possible to "sense" at that level. The level of aggregation of the most detailed acquired "data-burst image" was arbitrarily assigned a value of unity, and only "zoom-out" was carried out with increasing levels of aggregation. The residual curvature value when further zoom-out could not be carried out is assumed to be a x-value of infinity. However, in practice, it represented a level when the entire "frame" was aggregated [32] to 9 (3x3) points. The curvature value at this aggregated extremum is called the "Residual Curvature" of the system. A non-zero value of residual curvature represents a measure of the energy exchange rate through the system. Due to the nature of our conservation assumption, only a magnitude of the residual curvature is obtained. It is both positive and negative, simultaneously, that requires any energy absorbed (by the system as a whole) to be also released within the time step under consideration, and vice versa.

Since we do not have any physical data more detailed than the captured level in our "databurst frame", we arbitrarily assume that the plot is symmetric to the left and right of x=1 and drops to zero at x=0.

We calculate $\kappa, \frac{1}{LTilda}, \frac{1}{L}$ at each aggregation levels. The aggregation level represents the number of points that were considered together as a unit for the specific calculation. From Fig. 1, It is attempted to extrapolate the Kappa graph backwards for lower values of x extending to zero by using the assumed symmetry condition. After this, the $\frac{1}{LTilda}$ line is extended backwards to intersect the mirror image (about y-axis) of the calculated Kappa line. The existence of a process zone associated with local instability is assumed. It is further assumed (Fig.1) that the process zone size instantly jumps to point B if it reaches point A, provided the y-value (or curvature) at B is lower than A. The log(x) value at intersection point A is converted to fractional number of points relevant to the aggregated frame under consideration, and is scaled by the actual frame under consideration to reach an estimate for the critical chain length. The intersection of $\frac{1}{LTilda}$ line provides an estimate for the long term critical chain length while intersection with the $\frac{1}{L}$ line provides a short term estimation of the critical chain length.

Prognosis of ACL Reconstruction: A Data Driven Approach

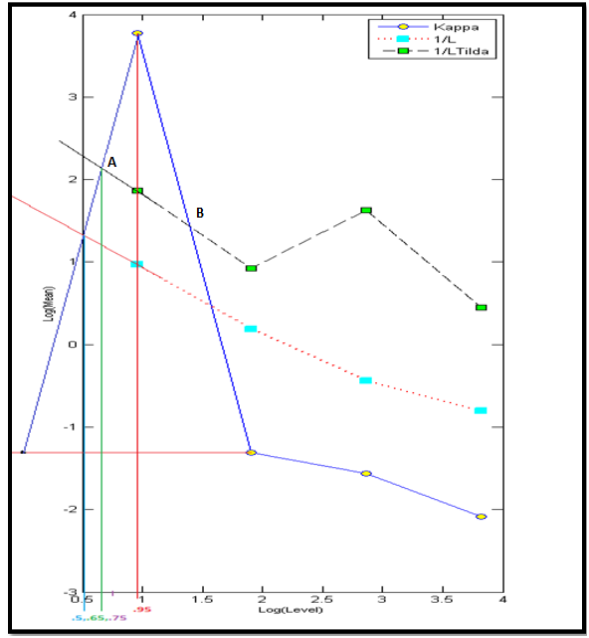

*Figure 1: Critical localization index calculation for Balloon Burst (online version in color)*

## V. STAIR ASCENT AND DESCENT EXPERIMENTS

### A. Experimental Observation and Data Collection

Individuals >1 year from unilateral ACLR and healthy controls 18-35 years old were recruited as research participants. This study was approved by the Institutional Review Board at Iowa State University. The post-ACLR group was on average 6 years from reconstruction surgery (range 2–18 years). The reconstruction type varied: 41% hamstring graft, 41% patellar tendon, 12% combination of cadaver and hamstring, and 6% cadaver. Most of the post-ACLR group sustained meniscal damage at time of ACL rupture (71%), and 59% of the ACL ruptures were considered non-contact in nature. After informed consent was obtained, age, height, and weight were recorded.

A three-step staircase was used. Kinematic data were collected using an 8-camera motion analysis system (Vicon Nexus). Kinetic data were recorded using two portable force platforms positioned on the first and second step of the stairs (AMTI). Kinematic data were collected at a sampling rate of 160 Hz, while force platform data were collected at a rate of 1600 Hz. Twenty-eight retro-reflective markers for static standing and 20 for dynamic movement were placed on the feet, calves, thighs, pelvis, and trunk. Participants performed three trials leading with each leg for a total of 12 stair ascent and stair descent trials.

Data were analyzed for the second step of stair ascent and descent. The stance phase was initiated at 20% BW and terminated when vertical ground reaction force dropped below 20% BW. Gait analysis was performed on the injured leg of the post-ACLR group and the right leg of the control group. Noise was reduced in kinematic and kinetic data using a fourth order, symmetric Butterworth filter with a cutoff frequency of 6 Hz. Using inverse dynamics, knee extension, external knee varus, and knee internal rotation moments were calculated during each stance phase and transformed to the distal segment coordinate system.

### B. Collected Data Files

The input files are in "xyzm" format. Individual subjects represented conservative and isolated systems. Each "data-burst" consisted of 81 time sequences of data, which were assumed to fully interact with each other. Each observation point has 4 dimensions or 4 types of information (xyzm: knee extension moment, knee varus moment, knee rotation moment and mechanical energy (kinetic energy + change in potential energy during data-burst)). The proposed DDP algorithm reduces memory requirements by parsing such data, and utilizing only frames (or data-bursts) at $n\Delta t$ (n = 1, 2, 3, etc.) intervals. Thus, past memory over only $n\Delta t$ is needed at any time instant. The proposed DDP algorithm can function well with n = 1, while higher values of n extends the prediction horizon. Of course, the implicit assumption in the prognosis scheme remains that the dimensionless length scales calculated over previous $n\Delta t$ remains valid over future $n\Delta t$ steps. This limits the upper bound of n or the prediction horizon that can be used at an instant.

## VI. MODEL BASED PROGNOSIS

### A. Path Dependency Index

Since the data size grows very rapidly, the first objective of our analysis is to make a prognosis using current observation, and hold a minimal number of "data-bursts" or frames in memory. As a first attempt, only two frames (current frame and its immediate predecessor) are used at each time instant.

First, the local Borda Count and Objective Rank are calculated at each of the observation point, at each instant of time. Using two consecutive time instances, the dimensionless length scales are calculated at each observation point. This allows calculation of the local curvature at each observation point. The local curvature at a time step is compared to the threshold value or critical curvature (denoted by inverse of length scale), and a Path Dependency Index (PDI) is calculated (Table 1 [31, 33]). A PDI greater than or equal to 5 indicates a local instability with potential to transcend to global scale.

TABLE 1. PATH DEPENDEMCY INDEX CATEGORIES

| Path Dependency Index Categories | |
|---|---|
| *Categories of classification* | *Meaning of categories* |
| Category 1 | Full Stability : $\frac{abs(Kappa)}{abs(KappaLong)} < 1$ and $\frac{abs(Kappa)}{abs(KappaShort)} < 1$ |
| Category 2 | Short term instability but long term stability: $\frac{abs(Kappa)}{abs(KappaShort)} > 1$ |
| Category 3 | Short term stability but long term instability: $\frac{abs(Kappa)}{abs(KappaLong)} > 1$ |
| Category 4 | Conditional Stability: Possibility exists for controlling local instability by altering mode mixity of dialational and distortion modes. |
| Category 5 | Both short term and long term instability. Control by alteration of mode mixity is not possible. Instability in single dimension. |
| Category 6 | Both short and long term instability in more than one dimension. |
| Category 7 | Both short and long term instability in more than two dimensions |
| Category 8 | PDI>5, Energy Exchange Channel(s) formed with Chain Length > short term critical chain length, |
| Category 9 | PDI>5, Energy Exchange Channel(s) formed with Chain Length > long term critical chain length |

Next, we follow the "Zoom Out" procedure to calculate the residual curvature, and identify the critical "chain length" or length of the energy exchange pathway needed for local instabilities to transcend to global scales.

However, the existence of a chain length greater than the minimum or critical length only constitutes a necessary condition for such transcendation. The energy exchange rate through such a pathway must also exceed a threshold value to meet the sufficiency condition for transcendation of local instabilities to a global scale. The critical energy release rate is a constitutive property that can be normally determined accurately only with careful off-line testing. Instead of such off-line testing, we utilize the fact that whenever local instabilities transition to global scales, almost all of the energy stored along such a pathway gets released (transformed to another form). Thus, the energy exchange rate rises, and then falls very rapidly to near zero during such transcendation phenomena. We utilize this rapid change in energy release rate in our prognosis scheme, and a trigger is initiated whenever the dimensionless energy release rate drops by more than 80% within a single time step. The choice of 80% is arbitrary in this case, based on inherent noise floor in our experimental and computational procedures. Together, the existence of: (i) greater than critical chain length, and (ii) greater than 80% drop in energy release rate over a single time step, constitute a positive reading for the GTI, and GTI is set >0. We continue the calculation of PDI and GTI for every time step of acquired data to prognosticate about the balloon burst phenomenon. When PDI >5 and GTI >0, imminent instability is indicated.

Based on PDI and GTI estimations, none of the subjects showed imminent instability. Due to limitations in data collection procedure and human subject involvement, only two "data-burst" frames were available. The calculation of residual curvature itself required two such frames. Hence, the drop in Residual Curvature value could not be meaningfully calculated in this study.

### B. Residual Curvature

Using residual curvature values from the multi-physics model, median values were determined for the 16 simulations in each dimension (knee extension moment, external knee varus moment, knee internal rotation moment, mechanical energy) and for the 64 simulations across dimensions. The percentage of simulations above a residual curvature of 3.0 (approximately 10 times the overall median value) was calculated for each dimension and across dimensions to assess the occurrence of excessive imbalances between knee joint moments and whole body movements. Dimensions with the highest percentage of residual values above 3.0 were broken down into distributions of 0-0.3, 0.3-1.5, 1.5-3.0, and >3.0. In addition, the residual curvature calculations in each dimension consisted of 16 roots ($2^4$, when 2 is the order of potential and 4 is the number of dimensions). The distribution of these roots was investigated in detail.

Fourteen individuals post-ACLR (8 female/6 male, age 24±2 year, height 1.76±0.13 m, mass 77±13 kg) and twelve healthy controls (7 female/5 male, age 26±4 year, height 1.69±0.14 m, mass 67±12 kg) participated in this study.

The post-ACLR group displayed higher median residual curvature (RC) values for knee extension moment, external knee varus moment, knee internal rotation moment, and combined dimensions during stair ascent (Table 1). In contrast, the control group had a higher median RC value for the kinetic energy dimension, although these values were much smaller than the other dimensions. The post-ACLR group displayed higher median RC values for external knee varus moment, knee internal rotation moment, and combined dimensions during stair descent (Table 2). Median RC values for knee extension moment and kinetic energy dimensions were within 5% when comparing groups. The greatest differences in median RC values were increases for the post-ACLR group: 37% in external knee varus moment dimension during stair descent, 32% in external knee varus moment dimension during stair ascent, and 30% in knee internal rotation moment dimension during stair descent.

Since the median RC values in various dimensions in the control group were about 0.3, we set a 10X threshold, and identified percentage population in both control and post-ACL group that displayed RC values greater than 3.0 in various dimensions. The post-ACLR group displayed a higher percentage of population with a RC value >3.0 for knee internal rotation moment and combined dimensions during stair ascent (Table 3). Percentage of RC values >3.0 for other dimensions were within 1% when comparing groups. The post-ACLR group displayed a higher percentage of RC values >3.0 for all dimensions during stair descent (Table 4). Greatest differences in percentage of RC values >3.0 were an increases in the post-ACLR group from 6.77% to 12.95% in the external knee varus moment dimension and from 6.25% to 11.61% in the knee internal rotation moment dimension for stair descent. Distributions of RC values for external knee varus moment and knee internal rotation moment dimensions during stair descent are shown in Figures 2 and 3.

TABLE 2. MEDIAN RESIDUAL CURVATURE VALUES IN STAIR ASCENT

| Dimension | Control | Post-ACL | % Change |
|---|---|---|---|
| Knee Extension | 0.470 | 0.508 | +8.00% |
| Knee Varus | 0.364 | 0.480 | +32.05% |
| Knee Rotation | 0.422 | 0.478 | +13.32% |
| Mechanical Energy | 0.036 | 0.028 | -22.59% |
| Combined Dimensions | 0.300 | 0.337 | +12.19% |

TABLE 3. MEDIAN RESIDUAL CURVATURE VALUES IN STAIR DESCENT

| Dimension | Control | Post-ACL | % Change |
|---|---|---|---|
| Knee Extension | 0.635 | 0.613 | -3.40% |
| Knee Varus | 0.405 | 0.557 | +37.48% |
| Knee Rotation | 0.444 | 0.576 | +29.68% |
| Mechanical Energy | 0.068 | 0.070 | +2.78% |
| Combined Dimensions | 0.305 | 0.374 | +22.79% |

TABLE 4. MEDIAN RESIDUAL CURVATURE > 3.0 IN STAIR ASCENT

| Dimension | Control | Post-ACL |
|---|---|---|
| Knee Extension | 6.25% | 6.25% |
| Knee Varus | 3.12% | 4.02% |
| Knee Rotation | 1.56% | 5.80% |
| Mechanical Energy | 1.04% | 0.89% |
| Combined Dimensions | 2.99% | 4.24% |

TABLE 5. MEDIAN RESIDUAL CURVATURE > 3.0 IN STAIR DESCENT

Prognosis of ACL Reconstruction: A Data Driven Approach

| Dimension | Control | Post-ACL |
|---|---|---|
| **Knee Extension** | 8.33% | 9.82% |
| **Knee Varus** | 6.77% | 12.95% |
| **Knee Rotation** | 6.25% | 11.61% |
| **Mechanical Energy** | 3.13% | 6.25% |
| **Combined Dimensions** | 6.12% | 10.15% |

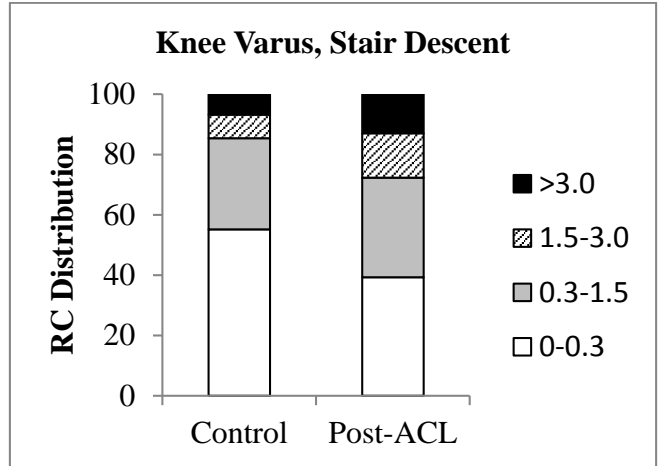

*Figure 2: Residual Curvature Distribution in Knee Varus in Stair Descent*

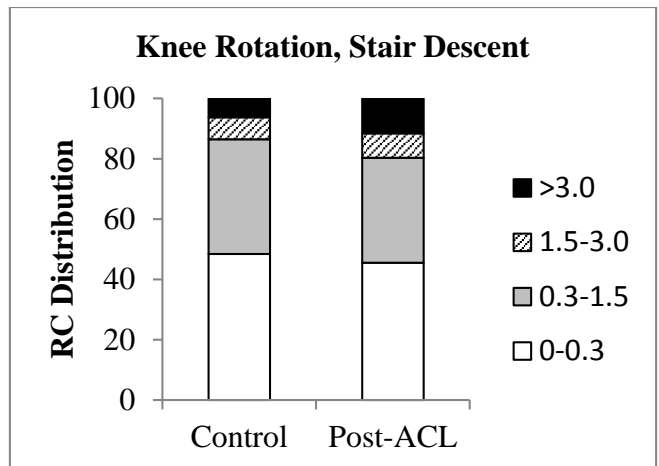

*Figure 3: Residual Curvature Distribution in Knee Rotation in Stair Descent*

Modulation amplitudes (within a databurst frame) in RC provide a measure of "jerkiness" in motion in respective dimensions, and may also be used as an indicator of the state of recovery post-ACLR. Distribution of individual roots calculated by the DDP algorithm facilitates a glimpse into such characteristics. Sample RC modulation amplitudes are shown in Figs. 4-7. Subjects 1-15 are controls, while 30-46 are post-ACLR population. The box extends from $25^{th}$ percentile to $75^{th}$ percentile, while the lines around it extends between +/- 2.7*standard deviations. Finally the outlier points are ploted as '+'. Such observations can provide individualized assessments for post-ACLR recovery. Figs. 4-5 shows RC modulations in stair ascent for knee extension and knee rotation respectively. The box-plots show that Subjects 35 and 37 display relatively high RC modulations in knee extension in stair ascent, which may be indicative of jerky motions in those dimensions. In Fig. 5, Subjects 30 and 39 display similar high RC modulations in knee rotation dimension, indicating jerky motions. Figs. 6-7 shows RC modulations in knee extension and knee rotation for stair descent. Here, we have a surprising result that Subject 11 (a control) displays very high RC modulation. Subjects 37 & 38 also show high RC modulation in knee rotation in stair descent.

Prognosis of ACL Reconstruction: A Data Driven Approach

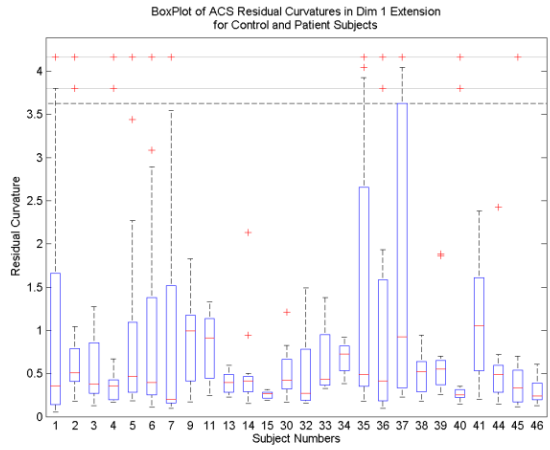

*Figure 4: Boxplots for Residual Curvature Distribution in Knee Extension in Stair Ascent (online version in color)*

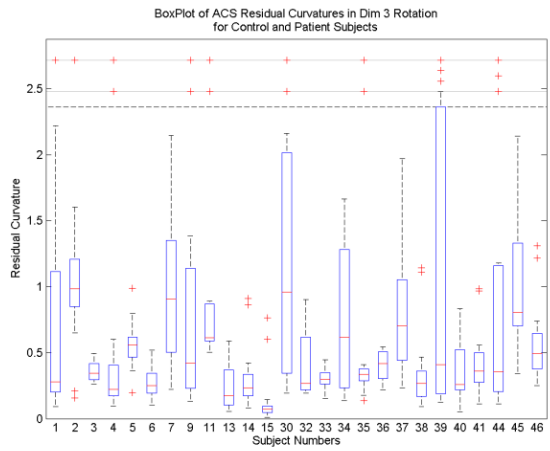

*Figure 5: Boxplots for Residual Curvature Distribution in Knee Rotation in Stair Ascent (online version in color)*

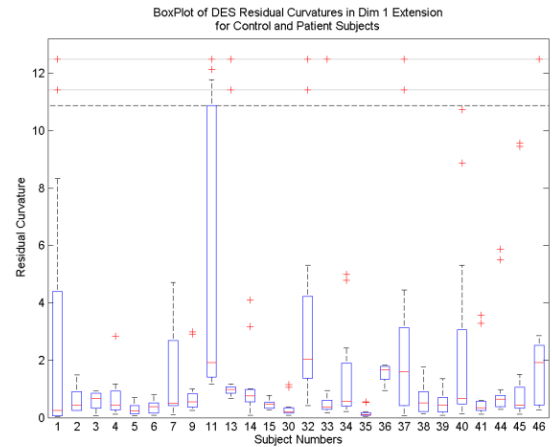

*Figure 6: Boxplots for Residual Curvature Distribution in Knee Extension in Stair Descent (online version in color)*

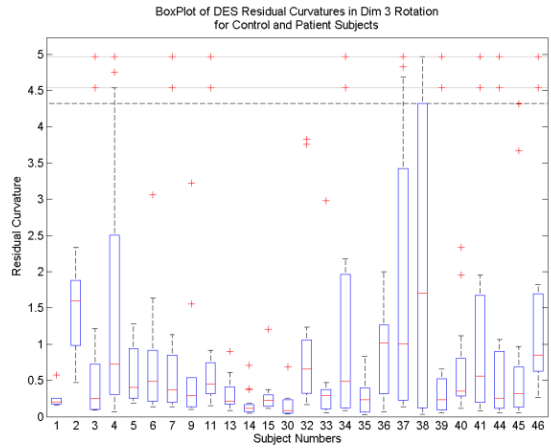

*Figure 7: Boxplots for Residual Curvature Distribution in Knee Rotation in Stair Descent (online version in color)*

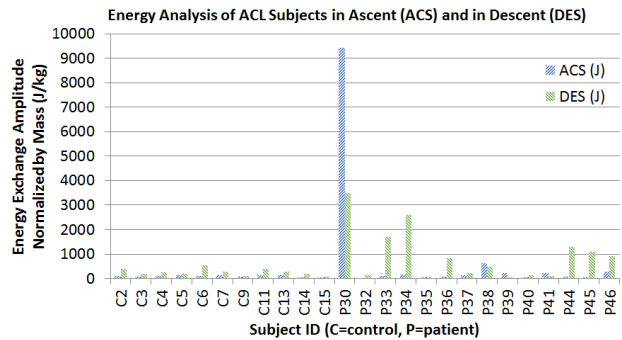

*Figure 8: Energy Exchange Amplitude, normalized by mass, of Patients and Controls for Stair Ascent and Stair Descent (online version in color)*

## VII. DISCUSSION AND CONCLUSION

A DDP algorithm suitable for handling multi-scale and multi-physics problems is applied to assessment of recovery after reconstructive surgery following ACL injury. Residual Curvature (RC) provides a measure of energy flowing through each dimension (knee extension, varus, rotation, as well as mechanical energy).

Stair ascent and stair descent are more difficult movements than walking, but present different mechanical challenges. Stair ascent is considered a strength challenge as body weight must be lifted against gravity, while stair descent is considered a control challenge as body weight must be lowered without losing balance. The multi-physics model provides evidence that stair descent is more challenging from an energy balance perspective, as median RC values and percent populations displaying RC >3.0 were greater than stair ascent. RC values for the kinetic energy dimension were much smaller than for the knee joint moment dimensions and will not be discussed further.

Knee extension moments are driven by a net balance between forces generated by the quadriceps, hamstring, and gastrocnemius musculature. Researchers have hypothesized that knee OA development post-ACLR may be due to quadriceps weakness and/or excessive co-contraction between muscles crossing the knee joint. Knee extension moment median RC values were highest for stair ascent/descent and percentage simulations >3.0 were highest for stair ascent. However, results were relatively similar when comparing post-ACLR and control groups. The multi-physics model indicates that knee extension moment generation is challenging from an energy balance perspective, but it is not a differentiator between post-ACLR and control groups.

Knee rotation moments are a complex balance that involves internal/external tibia/femur bone rotation and cruciate/collateral ligament resistance. The post-ACLR group displayed higher median RC values and percentage simulations >3.0 for knee internal rotation moments during stair ascent and descent. Energy imbalance in knee rotation moments is of concern in that it may indicate a higher probability for increased torsion of the knee joint, with resulting increased strains in the repaired ACL and/or increased shear forces in cartilage/menisci. Knee rotation moments have received relatively little attention in post-ACLR studies, and the multi-physics model indicates that this is a parameter of interest.

Prognosis of ACL Reconstruction: A Data Driven Approach

External knee varus moments are primarily resisted by knee joint cartilage/menisci. The post-ACLR group displayed higher median RC values and percentage simulations >3.0 for knee varus moments during stair ascent and descent. Energy imbalance in knee varus moments is of concern since it may indicate higher probability for increased compression of medial knee joint compartment cartilage/menisci, which is associated with progression of knee OA. Previous post-ACLR studies have not shown a consistent pattern of increased knee varus moments prior to advanced knee OA. In addition to median values of RC, the RC modulation amplitudes provide an indication of jerkiness in motion in the respective dimension.

Using energy balance, the multi-physics model may provide an analytical means for early detection of knee OA. Fig. 8 shows histograms of mass (or body weight) normalized energy exchange amplitudes (calculated as scalar or dot products of Residual Curvature Vector with the movement of center of mass) for different subjects. A relatively high energy exchange amplitude to achieve same stair ascend or descend task may be taken as an indication that the individual's knee is "working hard" to achieve the same task. By this measure, the subject P30 is working much harder in both ascend and descend. Also, P33 and P34 are displaying higher energy exchange amplitudes in stair descend. A longitudinal study is planned to further investigate implications of these insights.

The proposed DDP algorithm obviates the need for a priori off-line testing or "training" to extract the transitional statistics of the system. Instead of attempting to develop a capability for predicting system response under general conditions, it utilizes only on-line available data and short term memory (minimum two time frames are needed) to develop a prognosticator specializing in the "current" situation or "problem at hand", and makes a prediction regarding incipient instability over a relatively short prediction horizon.

The reliance of the DDP prognosticator on only short term memory significantly reduces the requirement that huge time sequences of data be held or handled to infer transition statistics of the system. However, big data sets must still be handled to extend the prediction horizon or provide prognosis over multiple prediction horizons.

Greater number of data-burst frames may be used to stretch the prediction horizon. For RUL estimates, the best type of data for the proposed algorithm are those collected over short data-burst windows separated longitudinally over a much larger time span. The separation time for the windows can be varied compared to the span of the data collection window to facilitate prognosis over varied prediction horizons. However, such a scheme requires capability for both short-term and long-term memory.

Current implementation of the proposed DDP algorithm is slow due to limitations of computational capability – both in terms of processor speed as well as number of processors deployed in parallel. Work is in progress to improve both of those aspects to approach real time implementation of the proposed DDP prognosticator.

ACKNOWLEDGMENTS

This work is supported by NSF through grant numbers: CMMI 0900093 and CMMI 1100066. The authors gratefully acknowledge this support. Any opinions, conclusions or recommendations expressed are those of the authors and do not necessarily reflect views of the sponsoring agencies.

Prognosis of ACL Reconstruction: A Data Driven Approach